\documentclass[10pt,lettersize,journal]{IEEEtran}
\IEEEoverridecommandlockouts 
\usepackage{blkarray}                                      
\usepackage{algpseudocode}                                 
\usepackage{algorithm}
\usepackage{graphicx}                                      
\usepackage{amsmath}
\usepackage{amssymb}
\usepackage{amsfonts}
\usepackage{amsthm}
\usepackage[mathcal]{eucal}
\usepackage{mathrsfs}
\usepackage{booktabs}
\usepackage{enumerate}
\usepackage{multirow}
\usepackage[subrefformat=parens,farskip=0pt,justification=centering]{subfig}
\usepackage{color}
\usepackage{cite}                                          
\usepackage{comment}                                       
\usepackage{soul}                                          
\soulregister\cite7
\soulregister\ref7
\soulregister\pageref7
\usepackage{etoolbox}                                      
\usepackage{url}
\usepackage{nth}                                           
\usepackage{bm}                                            
\usepackage{courier}
\usepackage{balance}
\usepackage{threeparttable}
\usepackage{xcolor,colortbl}
\usepackage{footnote}
\usepackage{listings}
\usepackage{setspace}                                      

\usepackage{verbatim}
\usepackage[bookmarks=false]{hyperref}
\hypersetup{
    colorlinks = true,
    citecolor  = blue,
    linkcolor  = blue,
    urlcolor   = blue,
}
\usepackage{tikz}
\usetikzlibrary{patterns,snakes}
\usetikzlibrary{positioning,calc,fit,decorations.pathmorphing,shapes.geometric, shapes.gates.logic.US, calc}
\usetikzlibrary{arrows,arrows.meta,decorations.markings,shapes,shapes.arrows}
\usetikzlibrary{decorations,decorations.pathreplacing}
\usetikzlibrary{backgrounds}
\usepackage{filecontents}                                  
\usepackage{pgfplots}
\usepackage{pgfplotstable}
\usepackage{scalefnt}
\pgfplotsset{compat=newest}
\usepackage{caption}
\usepackage{cleveref}
\Crefformat{figure}{Fig.~#2#1#3}                           
\Crefname{subfigure}{Fig.}{Figs.}
\Crefname{figure}{Fig.}{Figs.}
\Crefformat{table}{TABLE~#2#1#3}                           
\usepackage[figuresright]{rotating}
\usepackage[frozencache,cachedir=./minted]{minted}
\setminted{escapeinside=@@, fontsize=\tiny, numbersep=8pt, breaklines=true, linenos=false}
\usepackage{dialogue}
\usepackage{tcolorbox}
\tcbuselibrary{minted}
\newtcblisting{purplebox}[2][]{listing only, listing engine=minted, minted options={escapeinside=@@, fontsize=\tiny, numbersep=8pt, breaklines=true, linenos=false}, colback=CUHKpurple!5!white,
colframe=CUHKpurple!75!black,fonttitle=\scshape\small,
colbacktitle=CUHKpurple!85!black,breakable,enhanced,
left=0.5mm,
attach boxed title to top left={xshift=3mm, yshift=-2mm},
title={#2},#1}
\newtcblisting{greenbox}[2][]{listing only, listing engine=minted, minted options={escapeinside=@@, fontsize=\tiny, numbersep=8pt, breaklines=true, linenos=false}, colback=CUHKgreen!5!white,
colframe=CUHKgreen!75!black,fonttitle=\scshape\small,
colbacktitle=CUHKgreen!85!black,breakable,enhanced,
left=0.5mm,
attach boxed title to top left={xshift=3mm, yshift=-2mm},
title={#2},#1}
\usepackage{dashrule}
\newtcblisting{bluebox}[2][]{listing only, listing engine=minted, minted options={escapeinside=@@, fontsize=\tiny, numbersep=8pt, breaklines=true, linenos=false}, colback=CUHKblue!5!white,
colframe=CUHKblue!75!black,fonttitle=\scshape\small,
colbacktitle=CUHKblue!85!black,breakable,enhanced,
left=0.5mm,
attach boxed title to top left={xshift=3mm, yshift=-2mm},
title={#2},#1}
\usepackage{dashrule}

\definecolor{CUHKorange}{RGB}{244,106,18} 
\definecolor{CUHKblue}{RGB}{0,111,190}    
\definecolor{CUHKgreen}{RGB}{0,127,128}   
\definecolor{CUHKred}{RGB}{228,46,36}     
\definecolor{CUHKyellow}{RGB}{198,148,34} 
\definecolor{CUHKdark}{RGB}{114,44,114}   
\definecolor{CUHKmiddle}{RGB}{144,44,144} 
\definecolor{CUHKlight}{RGB}{167,44,167} 
\definecolor{CUHKpurple}{RGB}{117,15,109}
\definecolor{CUHKgold}{RGB}{221,163,0}
\definecolor{CUHKribbon}{RGB}{244,223,176}
\definecolor{CUHKblack}{RGB}{34,24,21}
\definecolor{lightgray}{RGB}{230, 230, 230}

\renewcommand{\bf}[1]{\textbf{#1}}
\renewcommand{\tt}[1]{\texttt{#1}}
\renewcommand{\vec}[1]{\boldsymbol{#1}}    

\newcommand{\minisection}[1]{\vspace{.1in}\noindent{\textbf{#1}}}

\usepackage{tcolorbox}
\tcbuselibrary{skins,breakable}
    {\endtcolorbox}

\usepackage[top=0.75in,bottom=0.80in,left=0.58in,right=0.58in]{geometry}
\newcommand{\subparagraph}{}
\usepackage{titlesec}
\titlespacing*{\section}{0pt}{1.8ex plus .2ex minus .2ex}{0.4ex plus .2ex}
\titlespacing*{\subsection}{0pt}{1.0ex plus .2ex minus .2ex}{0.2ex plus .2ex}
\crefname{mytheorem}{Theorem}{Theorems}
\crefname{mylemma}{Lemma}{Lemmas}
\crefname{myclaim}{Claim}{Claims}
\crefname{myproperty}{Property}{Properties}
\crefname{mycorollary}{Corollary}{Corollaries}

\algrenewcommand\textproc{\texttt}

\makeatletter
\let\OldStatex\Statex
\renewcommand{\Statex}[1][3]{%
  \setlength\@tempdima{\algorithmicindent}%
  \OldStatex\hskip\dimexpr#1\@tempdima\relax
}
\makeatother

\RequirePackage[normalem]{ulem} 
\RequirePackage{color}\definecolor{RED}{rgb}{1,0,0}\definecolor{BLUE}{rgb}{0,0,1} 

\usepackage{times}
\usepackage{enumitem}
\usepackage{enumerate}
\usepackage{listings}
\graphicspath{{./figs/}}

\begin{document}
\date{}

\title{
    ChatEDA: A Large Language Model Powered Autonomous Agent for EDA
}

\author{
    Haoyuan Wu$^{\dagger}$,     \ 
    Zhuolun He$^{\dagger}$\thanks{$\dagger$ Equal contributors},     \ 
    Xinyun Zhang,   \ 
    Xufeng Yao,     \ 
    Su Zheng,       \ 
    Haisheng Zheng, \ 
    Bei Yu
    \thanks{
        This work is supported in part by The Research Grants Council of Hong Kong SAR (Project No.~CUHK14210723).
        (\textit{Corresponding author: Bei Yu})
    }
    \thanks{Haoyuan Wu and Haisheng Zheng are with Shanghai Artificial Intelligence Lab, Shanghai, China.}
    \thanks{Zhuolun He, Xinyun Zhang, Xufeng Yao, Su Zheng, and Bei Yu are with the Department of Computer Science and Engineering, The Chinese University of Hong Kong, NT, Hong Kong SAR.}
}

\maketitle

\begin{abstract}

The integration of a complex set of Electronic Design Automation (EDA) tools to enhance interoperability is a critical concern for circuit designers.
Recent advancements in large language models (LLMs) have showcased their exceptional capabilities in natural language processing and comprehension, offering a novel approach to interfacing with EDA tools. 
This research paper introduces ChatEDA, an autonomous agent for EDA empowered by an LLM, AutoMage, complemented by EDA tools serving as executors.
ChatEDA streamlines the design flow from the Register-Transfer Level (RTL) to the Graphic Data System Version II (GDSII) by effectively managing task decomposition, script generation, and task execution.
Through comprehensive experimental evaluations, ChatEDA has demonstrated its proficiency in handling diverse requirements, and our fine-tuned AutoMage model has exhibited superior performance compared to GPT-4 and other similar LLMs.
\end{abstract}


\section{Introduction}
\label{sec:intro}


\IEEEPARstart{E}{lectronic} Design Automation (EDA) encompasses a crucial set of software tools utilized for circuit design, analysis, and verification.
These tools are organized within a complex design flow, featuring intricate programming interfaces.
Notably, advanced RTL-to-GDSII design platforms like OpenROAD~\cite{ajayi2019openroad} and iEDA~\cite{li2023ieda} consist of numerous procedures and adjustable parameters.
Commercial tools, with their extensive functionalities and options, offer even more comprehensive capabilities.
Circuit design engineers employ these tools iteratively to achieve their design objectives, often resorting to custom scripts for specific operations.
Conventionally, scripting languages such as TCL have been the \emph{de facto} means of interacting with EDA tools~\cite{chen2001scripting}, which is tedious and prone to errors.
Experienced design teams often adopt tools from different vendors, greatly increasing the difficulty in creating and maintaining such scripts. 

Recently, the field of natural language processing (NLP) has undergone a revolutionary transformation with the emergence of large language models (LLMs) such as GPT-3 \cite{brown2020language}, GPT-4 \cite{openai2023gpt4}, Claude2 \cite{2023claude2}, and Llama \cite{touvron2023llama, touvron2023llama2}.
Extensive training on large corpora enables large language models to acquire emergent abilities \cite{wei2022emergent} by learning intricate patterns and relationships in language.
This allows these models to demonstrate remarkable accuracy and fluency in a variety of NLP tasks, such as natural language understanding and generation.
To leverage their potential in specialized domains \cite{wang2022self, cui2023chatlaw, li2023chatdoctor}, instruction tuning \cite{wei2022finetuned} fine-tunes LLMs with domain-specific corpora, resulting in remarkable performance on these specialized domains.
Specifically, Vicuna \cite{zheng2023judging}, Guanaco \cite{dettmers2023qlora}, and Orca \cite{mukherjee2023orca}, have applied instruction tuning to train LLMs, making use of the outputs produced by the GPT, and thereby achieving significant outcomes

\begin{figure}[tb!] 
    \centering
    \includegraphics[width=0.86\linewidth]{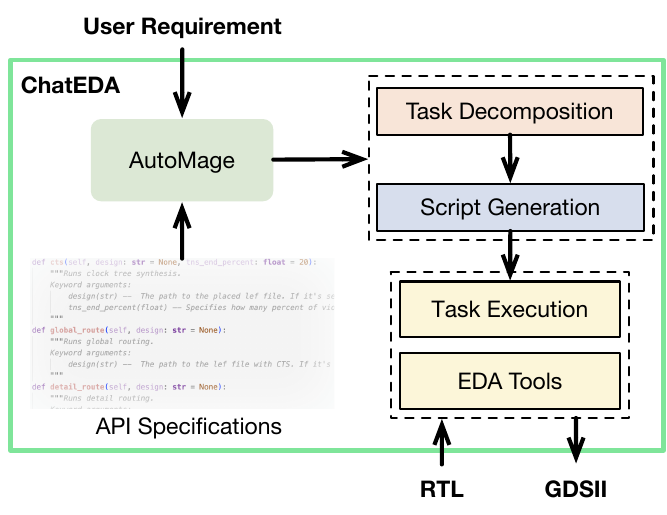}
    \caption{Overview of AutoMage powered ChatEDA. With AutoMage as the controller and EDA tools as the executors, the workflow consists of three stages: 1) Task Decomposition; 2) Script Generation; 3) Task Execution.} 
    \label{fig:chatEDA}
\end{figure}

Furthermore, scholars have initiated exploration into the incorporation of tools or models into LLMs.
Toolformer \cite{schick2023toolformer}, a groundbreaking methodology, integrates external API tags into text sequences, thus facilitating LLMs to connect with external tools.
This tool utilization, coupled with the capacity for logical reasoning, broadens the LLM's potential as a robust general problem solver.
Several proof-of-concept demonstrations, including AutoGPT \cite{2023autogpt} and BabyAGI \cite{2023babyagi} serve as motivational illustrations.
The current implementation of LLMs in toolchain automation predominantly relies on generic LLMs without specific fine-tuning. 
However, such LLMs, lacking bespoke fine-tuning, are unable to consistently meet performance standards tailored to users' specific requirements \cite{patil2023gorilla}. 
Particularly in the EDA domain, LLMs exhibit limited familiarity with EDA toolchains, leading to frequent errors during the tool usage process. 
In this paper, we introduce the expert EDA LLMs, the AutoMage series, which have been optimized for proficiency with EDA tools, thereby enhancing the stability and reliability of the automation of EDA workflows.

In this work, we propose ChatEDA, an expert LLM system designed to generate code for manipulating EDA tools based on natural language instructions.
To be more specific, as illustrated in \Cref{fig:chatEDA}, ChatEDA is an LLM-driven autonomous agent system for EDA, functioning as the agent's intellectual hub, responding to human instructions and manipulating the EDA tools via APIs to deliver autonomous Register-Transfer Level (RTL) to Graphic Data System Version II (GDSII) capabilities without necessitating any code writing.
To guarantee the performance, we utilize the AutoMage series (AutoMage and AutoMage2) as the control unit of the ChatEDA.
AutoMage is an expert LLM that specializes in the utilization of EDA tools, which is barely learned in open-source LLMs.
To further enhance AutoMage’s abilities in real-world environments, we propose the upgraded version of AutoMage, AutoMage2.
Our contributions are listed as follows:
\begin{itemize}
    \item ChatEDA, the first LLM-powered EDA interfacing framework and methodology;
    \item AutoMage series (AutoMage and AutoMage2) are fined-tuned based LLMs and purpose-built to enhance the capabilities of ChatEDA; and 
    \item Comprehensive evaluations using ChatEDA-Bench to show the superior performance of AutoMage series, surpassing GPT-4 and other well-known LLMs in various tasks.
\end{itemize}

The remainder of this article is organized as follows. 
\Cref{sec:prelim} discusses the preliminaries, including a brief illustration of the generative pre-trained language model, low-rank adaptation of LLMs, in-context learning ability of LLMs, and block-wise k-bit quantization technique for efficient training. 
Our proposed LLM-powered framework for EDA will be explained in \Cref{sec:chateda}.
AutoMage and its upgraded version AutoMage2 will be elaborated in \Cref{sec:automage} and \Cref{sec:automage2} separately.
\Cref{sec:exp} demonstrates our experiment setup, evaluation of our methods, quantitative comparisons, and some case studies, followed by a discussion about limitations and future work in \Cref{sec:discussion} and a conclusion in \Cref{sec:conclu}.

\section{Preliminaries}
\label{sec:prelim}

\subsection{Generative Pre-Trained Language Model}
Generative pre-trained language models (GPLMs) stand at the pinnacle of natural language processing advancements. 
Unlike conventional models such as BERT \cite{devlin2018bert} and XLNet \cite{yang2019xlnet}, which utilize an encoder-decoder architecture, GPLMs \cite{brown2020language,openai2023gpt4,2023claude2,touvron2023llama,touvron2023llama2} employ a neural network structure exclusively comprising decoder blocks based on the transformer architecture. 
This exclusive decoder-only design offers several advantages over traditional encoder-decoder architectures \cite{tay2022ul2}.
By eschewing the encoder's compression of input into a singular vector, GPLMs adeptly capture long-range language dependencies, leading to the synthesis of more coherent prose. 
The auto-regressive decoder forecasts each token considering its entire antecedent context, ensuring fluid and logical text generation.

GPLMs are trained through self-supervised learning on expansive corpora, cultivating broad language representations. 
In this phase, model parameters are refined to augment the likelihood of predicting ensuing tokens in sequences \cite{radford2018improving}. 
Such a strategy equips these models with a deep comprehension of linguistic intricacies. 
As training data swells, the proficiency of these models in text generation augments \cite{brown2020language}.

Significantly, while these models are not engineered for distinct downstream tasks, their broad knowledge garnered during pretraining paves the way for stellar performance across diverse tasks. 
This is achieved with minimal fine-tuning on scantily labeled datasets \cite{brown2020language}. 
Their pre-trained representations coupled with the decoder-centric design empower them to craft coherent, sensible, and fluid prose infused with reasoning. 
The blend of generative prowess and pre-acquired knowledge paves the path for imaginative text generation.

Prominent LLMs, including GPT-4 \cite{openai2023gpt4}, PaLM \cite{chowdhery2022palm, anil2023palm2}, and LLaMA \cite{touvron2023llama,touvron2023llama2}, epitomize GPLMs. 
They display unparalleled generalization and few-shot learning prowess, endorsing a plethora of text generation applications. 
In the context of this paper, AutoMage is fine-tuned based on GPLMs (Llama2), underlining the adaptability and potency to automate the RTL-GDSII flow in real-world scenarios.

\subsection{Low-Rank Adaptation of LLMs}
LLMs are characterized by their vast number of parameters, making full fine-tuning of these parameters during training impractical. 
An efficient alternative is low-rank adaptation (LoRA) \cite{hu2021lora}, a technique that involves preserving the pretrained model weights while introducing trainable low-rank decomposition matrices into each layer of the Transformer architecture. 
This method substantially decreases the number of trainable parameters for subsequent tasks.

The Transformer architecture comprises numerous fully connected layers that conduct matrix multiplications with full-rank weight matrices. 
Despite the complexity, pre-trained language models demonstrate a low intrinsic dimension, allowing them to learn efficiently even after random projection into a smaller subspace \cite{aghajanyan2021intrinsic}. Consequently, for a pre-trained weight matrix represented as $\vec{W} \in \mathbb{R}^{h\times d}$, it can be updated using a low-rank decomposition $\vec{W} + \Delta \vec{W} = \vec{W} + \vec{L}_{1}\vec{L}_{2}$,
where $\vec{L}_{1} \in \mathbb{R}^{h\times r}$, $\vec{L}_{2} \in \mathbb{R}^{r\times d}$, and the rank $r \ll \min(h, d)$.

During the training process, $\vec{W}$ remains fixed and does not receive gradient updates,
whereas the matrices $\vec{L}_{1}$ and $\vec{L}_{2}$ are endowed with trainable parameters. 
For the equation $\vec{y} = \vec{W} \vec{x}$, the modified forward pass is expressed as:
\begin{equation}
    \vec{y} = \vec{W} \vec{x} + \Delta \vec{W} \vec{x} = \vec{W} \vec{x} + \vec{L}_{1}\vec{L}_{2} \vec{x}.
    \label{Eq:LoRA}
\end{equation}
Initially, $\vec{L}_{2}$ is initialized with a random Gaussian distribution, and $\vec{L}_{1}$ is set to zero,
ensuring that $\Delta \vec{W} = \vec{L}_{1}\vec{L}_{2}$ is zero at the outset. 

During production deployment, $\vec{W} = \vec{W} + \vec{L}_{1}\vec{L}_{2}$ is explicitly computed and stored for regular inference. 
Importantly, this method incurs no additional latency when compared to a fully fine-tuned model, making it an efficient choice for practical applications.

In the context of this study, LoRA is applied to streamline the fine-tuning process for Llama2. By preserving the pre-trained weights and solely updating the low-rank matrices, the number of trainable parameters is significantly reduced, enhancing the model's efficiency and applicability in real-world scenarios.

\subsection{In-Context Learning}
\label{sec:icl}
The concept of in-context learning (ICL) \cite{min2022metaicl} exemplifies the remarkable ability of LLMs to execute downstream tasks effectively by conditioning on the in-context prompt containing a limited number of input-output examples, all without explicit fine-tuning. 
For instance, when presented with a task like predicting nationalities, a prompt featuring sample input names and their respective nationalities,
such as ``Albert Einstein was German. Mahatma Gandhi was Indian. Marie Curie was \_\_\_\_'',
allows LLMs to correctly fill in the blank with the appropriate nationality.

The phenomenon of ICL arises from the presence of long-range coherence in pre-training documents. 
During the pre-training phase, LLMs are compelled to deduce the latent concept denoted by $\theta$ across multiple sentences, ensuring coherent continuations. When provided with the in-context prompt, denoted as $x_{ic}$, ICL manifests when LLMs deduce shared concepts within $x_{ic}$ to make predictions denoted as $x_{o}$.

Assuming that LLMs precisely capture the pre-train distribution $p$ with adequate data and expressivity \cite{xie2021explanation}, ICL involves characterizing the conditional distribution of completions given in-context prompt, denoted as $p(x_{o}|x_{ic})$, under $p$. 
This is the posterior predictive distribution, which marginalizes out latent concepts as follows:
\begin{equation}
    p(x_{o}|x_{ic}) = \int_{\theta} p(x_{o}|\theta,x_{ic}) p(\theta|x_{ic}) d(\theta).
    \label{Eq:ICL}
\end{equation}
In scenarios where $p(\theta|x_{ic})$ focuses on the concepts within $x_{ic}$ with more input-output examples, LLMs learn through marginalization by effectively ``selecting'' the concept of input-output examples from $x_{ic}$.

ICL \cite{dong2022survey} presents an efficient and adaptable method to leverage the knowledge and capabilities embedded within extensively pre-trained language models. 
It stands as a promising paradigm enabling LLMs to learn from a minimal set of examples during inference. 
In our research, we employ ICL for self instruction and the collection of instruction datasets to fine-tune Llama2.

\subsection{Block-wise k-bit Quantization}
Quantization \cite{frantar2022gptq}, a process of discretizing input from a high-information representation to a lower-information one, involves converting data types with more bits into those with fewer bits, such as transitioning from 32-bit floats to 8-bit integers. 
Block-wise k-bit quantization ensures optimal utilization of the low-bit data type's entire range. 
This approach normalizes the input data type within the target data type's range using the absolute maximum of the input elements, typically organized as a tensor.
For instance, quantizing a 32-bit Floating Point (FP32) tensor into an Int8 tensor within the range of $[-127, 127]$ can be expressed as:
\begin{equation}
    \begin{split}
        X^{\text{Int8}} &= \text{round}\left(\frac{127}{\text{absmax}(X^{\text{FP32}})}X^{\text{FP32}}\right) \\
        & = \text{round}(c^{\text{FP32}} \cdot X^{\text{FP32}}),
    \end{split}
    \label{Eq:quant}
\end{equation}
where $c$ represents the quantization constant. The inverse operation dequantization is defined as:
\begin{equation}
    \text{dequant}\left(c^{\text{FP32}}, X^{\text{Int8}}\right) = \frac{X^{\text{Int8}}}{c^{\text{FP32}}} = X^{\text{FP32}}.
    \label{Eq:dequant}
\end{equation}
The input tensor $X \in \mathbb{R}^{b\times h}$ is divided into $n$ contiguous blocks of size $B$ by flattening the tensor and segmenting the linear structure into $n = (b \times h)/B$ blocks. 
These blocks are independently quantized using \Cref{Eq:quant}, creating a quantized tensor and $n$ quantization constants $c_{i}$.

In the context of efficient fine-tuning of quantized LLMs, a strategy involving block-wise k-bit quantization is applied. 
To be more specific, the pretrained LLM is stored in a 4-bit datatype using \Cref{Eq:quant},
and then dequantized from 4-bit to 16-bit datatype for forward and backward pass computations using \Cref{Eq:dequant}.

\section{ChatEDA: LLM-powered Framework for EDA}
\label{sec:chateda}

ChatEDA, an LLM powered agent, is specifically designed for RTL-to-GDSII flow automation.
The main objective of ChatEDA is to understand and respond to user requirements in natural language.
In order to achieve this, ChatEDA is capable of breaking down complex user requirements into smaller, more manageable sub-tasks and subsequently utilizing appropriate EDA tools to address them.

\begin{figure*}[tb!] 
    \centering
    \includegraphics[width=0.928\linewidth]{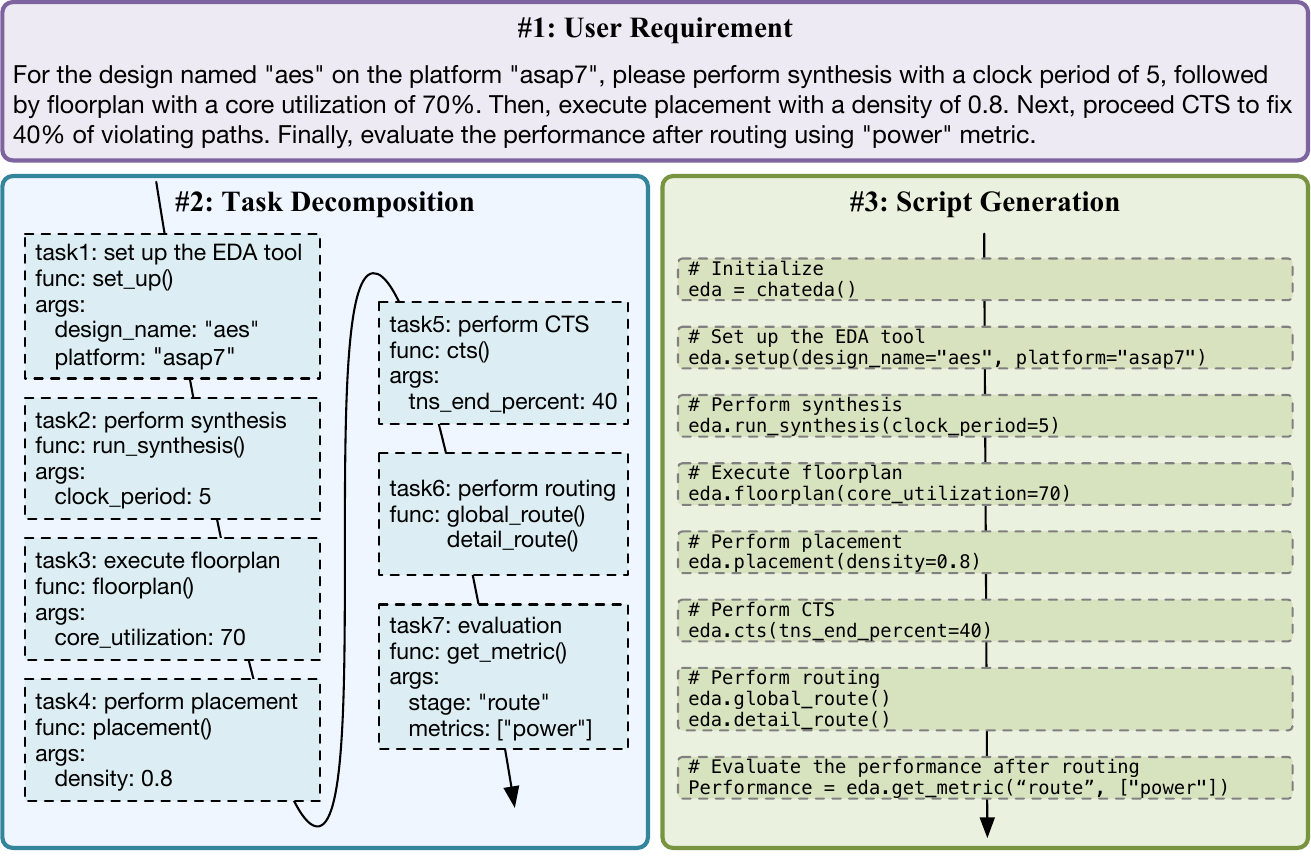}
    \caption{Language functions as a conduit enabling ChatEDA to integrate EDA tools for resolving complex EDA tasks. Within the framework, ChatEDA acts as the controller that harmonizes and orchestrates the collaboration among various tools. ChatEDA first formulates a task list derived from user requirements, subsequently generating scripts corresponding to these decomposed tasks.} 
    \label{fig:chatEDA-agent}
\end{figure*}

As illustrated in \Cref{fig:chatEDA}, AutoMage, an LLM fine-tuned with EDA expert knowledge, serves as the central processing unit of ChatEDA. 
After receiving a natural language requirement from the user, AutoMage first interprets the requirement and decomposes it into a set of sub-tasks, known as task decomposition.
Then, based on the decomposed smaller tasks and the specifications for the external tools, e.g., OpenROAD, AutoMage generates Python scripts for accomplishing these tasks. 
Ultimately, ChatEDA executes the generated script to get the final output for the user requirement.
We will detail the workflow and the training process of AutoMage in the following sections.

\subsection{Task Decomposition}

In the realm of automating the RTL-to-GDSII flow through EDA tools, numerous user requests often entail intricate intentions. A fundamental requirement lies in the agent's ability to comprehend these complex human natural language requests. Thanks to the robust capabilities of AutoMage, ChatEDA adeptly interprets these tasks based on human-defined specifications.

Considering the intricate nature of automating the RTL-to-GDSII flow, it becomes imperative to break down the overarching task into a series of manageable sub-tasks to achieve the desired outcome. Therefore, we introduce task decomposition as the primary stage of ChatEDA. In this phase, AutoMage assesses user requirements and dissects them into a sequence of structured tasks. \Cref{fig:chatEDA-agent} illustrates the task decomposition process. For instance, when faced with a convoluted and lengthy natural language request, ChatEDA employs AutoMage to break it down into a series of specific sub-tasks. These tasks, encompassing aspects like logic synthesis, floorplan, placement, and others, can then be efficiently handled through various EDA tools.

\subsection{Script Generation}

Upon the completion of the task decomposition phase, manageable sub-tasks are defined, facilitating the streamlined orchestration of the complex task. Each sub-task is executable through corresponding APIs within the EDA tools. Consequently, the need arises to craft a script that invokes these APIs for task execution. During the script generation phase, depicted in \Cref{fig:chatEDA-agent}, a structured text incorporating API specifications, user requirements, and the decomposed sub-tasks serves as input for AutoMage. Subsequently, AutoMage generates a Python script, ready for direct execution. This script enables the RTL-to-GDSII flow, promoting efficient architectural exploration, design space evaluation, early Quality of Results (QoR) estimation, and detailed physical design implementation.

\subsection{Task Execution}

Following script generation, ChatEDA executes the script using the Python interpreter, and the sub-tasks are then performed utilizing EDA tools.
According to the setup procedures, ChatEDA sets environment variables accordingly.
Then, it launches a subprocess that runs the tool (i.e., OpenROAD in our implementation) script executor.
Internally, the Python wrapper implements different functionalities by specifying relevant Tcl scripts or commands and running them by the tool script executor.
With its proven efficacy in both script generation and task execution, ChatEDA stands as a pivotal system in ensuring reliable automation of the RTL-to-GDSII flow.

\section{AutoMage: LLM-based Controller of ChatEDA}
\label{sec:automage}

\subsection{Base Model Selection}
\label{sec:model-arc} 
The AutoMage model is a fine-tuned version of Llama2 \cite{touvron2023llama2}, which is designed based on the standard Transformer model architecture in a decoder-only setup, meaning each timestep can only attend to itself and past timesteps.
It is worth noting that CodeLlama \cite{roziere2023code} models achieve strong performance in coding ability, which are incrementally pretrained on code resources based on Llama2 models.
However, during the process of incremental pretraining on code resources, it will lose a lot of general knowledge including EDA knowledge.
To understand user requirements on EDA tool usage, ChatEDA needs to have basic EDA knowledge.
As a result, we utilize Llama2 models for this research.

\subsection{Instruction Tuning for AutoMage.}
\label{sec:instruct-tuning}
\begin{figure}[tb!] 
    \centering
    \includegraphics[width=0.968\linewidth]{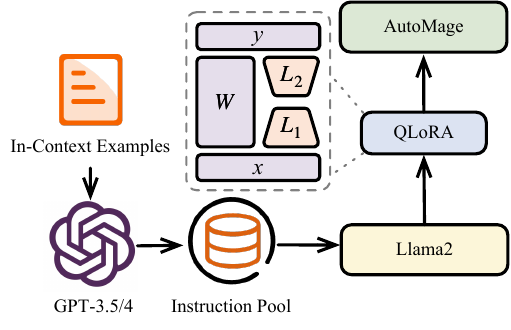}
    \caption{Overview of Instruction Tuning. 
    During the instruction tuning process, we use the self instruction paradigm to construct our instruction pool via GPT models. 
    Then we apply the QLoRA technique for efficient instruction fine-tuning.} 
    \label{fig:instruction-tuning}
\end{figure}

\begin{figure*}[tb!] 
    \centering
    \includegraphics[width=0.96\linewidth]{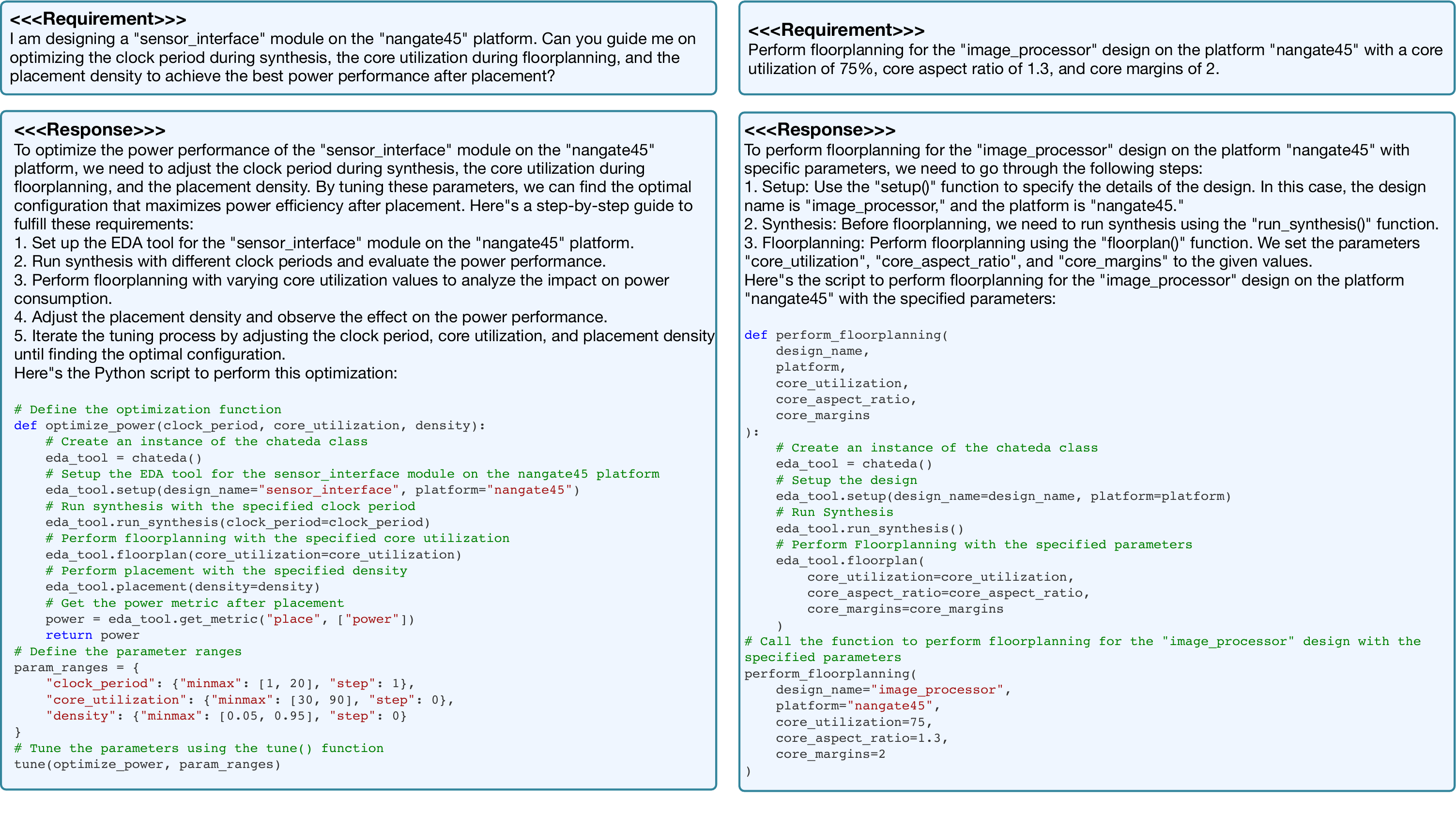}
    \caption{Examples of generated EDA tools instructions. Moreover, we also provide more examples in the repo \url{https://github.com/wuhy68/ChatEDAv1} for a better understanding of the generated dataset.} 
    \label{fig:dataset}
\end{figure*}

To guarantee the reliability of ChatEDA, knowing when and how to use the tools, which is determined by the LLM's capability, is of vital importance.
AutoMage is an expert LLM specializing in utilizing EDA tools, which are barely learned in open-source LLMs.
To integrate expert knowledge into LLMs, instruction tuning is an effective approach that enables LLMs to benefit from the pairing of domain-specific natural language descriptions and their corresponding responses.
Therefore, we incorporate instruction tuning to train AutoMage, the core controller of ChatEDA, based on the open-source LLMs (Llama2 \cite{touvron2023llama2}).
As shown in \Cref{fig:instruction-tuning}, the process of instruction tuning mainly includes three parts, self instruction, instruction collection, and instruction fine-tuning, which are detailed as follows.

\minisection{Self Instruction.}
\label{sec:self-instruct}
To enhance instruction tuning, it is imperative to gather high-quality instructions that can effectively educate Large Language Models (LLMs) on the utilization of EDA tools through APIs. 
Given the costliness of obtaining high-quality instructions, the self-instruction paradigm has gained significant traction in recent research endeavors \cite{ouyang2022training}. 
This paradigm revolves around the utilization of diverse in-context prompts, tailored to specific instances, to query GPT-3.5/4 and automatically generate additional instances from them.

The self-instruction paradigm is meticulously crafted, drawing inspiration from the ICL \cite{dong2022survey} capability of GPT-3.5/4. 
In this methodology, we provide explicit guidance to the model through APIs, incorporating necessary restrictions within the self-instruction prompts. 
Subsequently, we employ instances structured in the format $<$requirement, decomposition, script$>$ as in-context prompts, enabling precise alignment with GPT-3.5/4. 
According to the formulation outlined in \Cref{Eq:ICL}, GPT-3.5/4 generates instances that directly reference the instances provided within the in-context prompts. 
This approach not only refines the understanding of the given instructions but also enhances the model's ability to generate contextually relevant responses.

Herein lies the template for the self-instruction prompt, meticulously designed to maximize the efficiency of instructing LLMs in utilizing EDA tools via APIs. The template, marked by careful structuring and explicit directives, plays a pivotal role in shaping the in-context learning process of GPT-3.5/4, ensuring the acquisition of high-quality instructions and fostering the model's adeptness in the EDA domain.
\begin{bluebox}[minted language=text, minted options={fontsize=\scriptsize}]{Self-instruction Prompt}
You are an expert who can automate the RTL to GDSII flow by using a fully autonomous toolchain. You are invited to teach us how to create chips by using a fully autonomous toolchain for digital layout generation across die sizes, process nodes, and foundry options. 
As a result, @\textbf{you are now required to generate instances for EDA user requirements and the corresponding script generation by given APIs}@. I will provide you with the template of the instance, available Python APIs, and some instance examples to describe how to use APIs to generate scripts.
I need to learn how to use these APIs according to your generated instances.

The @\textbf{instance template}@ is as follows:
<Requirement>:
    This part is the requirement.
    You should try to make it diverse as if they are written by different people. 
    Try to customize needs to cover APIs as much as possible.
<Analysis>:
    This part is the analysis.
    According to the provided APIs, you are required to provide an analysis to illustrate how to fulfill user requirements.
<Script>
    This part is the script.
    You should only contain Python code with correct API calls.
</Script>

Here are the provided APIs of class 'chateda':
<<<API documents>>>

Here are some @\textbf{restrictions}@:
1. The instances should be in English.
2. Try not to repeat the verb or noun for each requirement and explanation to maximize diversity.
3. The requirements can be either an imperative sentence or a question. For example, you can combine questions with imperative requirements.
4. Your requirements and relative analyses should be professional and logical.
5. The analyses should be as detailed as possible. For example, you can think it step by step.
6. The requirements are required to be able to be completed using the provided APIs.
7. In the script, you are required to use named arguments in the Python code to allow passing arguments by specifying their names, independent of their positions.
8. The Python code in the scripts can be executed directly.
9. You can't skip any steps that need to be executed. The usual flow of chip designing goes like this in sequence (Note that each step follows can't be executed unless the previous step has been executed.): a. Setup; b. Synthesis; c. Floorplanning; d. Placement; e. Clock Tree Synthesis (CTS); f. Global Routing; g. Detailed Routing; h. Density Fill; i. Final Report;
10. Noted that if you are required to get a metric, You have to execute that process first.

Now I need you to teach me how to master these APIs according to the provided instances. Please generate more different instances according to the provided instance examples under the provided restrictions. 

Here are some @\textbf{instance examples}@:
<<<in-context instances>>>
\end{bluebox}

\minisection{Instruction Collection.} In adherence to the self-instruction paradigm, a meticulously crafted dataset of approximately 1500 instances was formulated for the explicit purpose of instruction tuning. 
It is crucial to note that both GPT-3.5 and GPT-4, while powerful, are not infallible, occasionally producing erroneous data. 
To counter this, a portion of the dataset was manually curated or refined to ensure accuracy and reliability. 
Specifically, we first automatically verified whether our designed OpenROAD API interface, an automated process, could correctly execute the generated code. 
In cases where the code was inexecutable, we modified it to ensure functionality. 
Subsequently, we manually evaluated whether the generated code satisfied the task requirements specified in the prompt. 
If the generated code failed to meet the requirements or exhibited an incorrect thought process, we manually edited it. 
For the entire dataset, we dedicated approximately two person-days to meticulously validate and refine the requirement-response pairs, ensuring quality and accuracy, resulting in a refined dataset of approximately 1500 training samples.

We provide some examples of the generated instruction dataset in \Cref{fig:dataset}.

\minisection{Instruction Fine-tuning.} In the process of fine-tuning, each instance consists of a requirement and a corresponding response. 
The response involves a detailed decomposition process and script. 
In this context, we view the prompt and requirement as crucial instructions that direct LLMs to generate accurate corresponding responses.
{During instruction fine-tuning, we utilize a commonly used prompt in the style of Alpaca \cite{alpaca}, and AutoMage is trained on approximately 1500 instances to learn how to utilize EDA tools.}

To ensure an appropriate length for the model sequences, requirements and responses extracted from the entire training set are concatenated. 
This concatenation is performed while employing a unique token to demarcate these segments effectively. 
Consequently, an auto-regressive objective is implemented, effectively nullifying the loss of tokens originating from the user requirement. 
This strategic approach confines the backpropagation process solely to the response tokens, enhancing the precision and efficiency of the fine-tuning process.

\subsection{Efficient Fine-tuning of Quantized LLMs}
\label{sec:qlora}
In addressing the critical necessity for swift training procedures, our study incorporates the QLoRA technique \cite{dettmers2023qlora}, aiming to expedite the fine-tuning process efficiently. 
QLoRA employs 4-bit Normalfloat (NF4) Quantization and Double Quantization methodologies, ensuring the attainment of high-quality 4-bit fine-tuning. 
This innovative approach is intricately coupled with paged optimizers, which serve the crucial purpose of mitigating memory spikes during gradient checkpointing, thus averting potential out-of-memory errors. 
The notable efficacy of QLoRA is pivotal in enabling us to ensure the seamless performance of instruction fine-tuning utilizing Large Language Models on a substantial scale, specifically 34B/70B models, a milestone traditionally hindered by memory overhead constraints.

To delve deeper into the components of QLoRA, it is imperative to elucidate the essence of the following fundamental components: NF4 quantization and double quantization.

\minisection{NF4 Quantization.} The NormalFloat (NF) data type, an extension of Quantile Quantization \cite{dettmers20218}, emerges as an information-theoretically optimal method, ensuring equal distribution of values within each quantization bin of the input tensor. 
Quantile quantization achieves this uniformity by estimating the quantile of the input tensor through the empirical cumulative distribution function.

\minisection{Double Quantization.} This sophisticated process involves quantizing the quantization constants to achieve additional memory savings. 
While precise 4-bit quantization \cite{dettmers2023case} necessitates a small block size, it inevitably results in significant memory overhead. 
Double quantization \cite{dettmers2023qlora}, a pivotal innovation, involves employing the quantization constants $c_2^{\text{FP32}}$ from the first quantization as inputs for a second quantization step. 
This process yields quantized quantization constants $c_2^{\text{FP8}}$ and a secondary set of quantization constants $c_1^{\text{FP32}}$. 
For the second quantization, we utilize 8-bit Floats with a block size of 256, ensuring optimal memory usage without compromising performance. 
Notably, to facilitate symmetric quantization, we center the positive $c_2^{\text{FP32}}$ values around zero after subtracting the mean before quantization.


\minisection{QLoRA.} Combining these aforementioned components, QLoRA enhances a linear projection within a transformer layer of LLMs through an additional factorized projection, thus contributing to the efficiency and effectiveness of fine-tuning processes in the realm of LLMs.
As shown in the dash box of \Cref{fig:instruction-tuning}, given a linear projection, $\vec{y}=\mathbf{W}\vec{x}$, the computation, based on \Cref{Eq:LoRA}, is as follows:
\begin{equation}
    \begin{aligned}
    \vec{y}^{\text{BF16}} ={}& \text{doubleDequant}(c_1^{\text{FP32}} \vec{x}^{\text{BF16}}, c_2^{\text{k-bit}}, \mathbf{W}^{\text{NF4}}) \\
    +{}&L^{\text{BF16}}_1L^{\text{BF16}}_2 \vec{x}^{\text{BF16}},
    \end{aligned}
\end{equation}
where $\mathbf{W}$ is original weights and $L_1$ and $L_2$ are additional QLoRA weights.
The final weights can be combined without extra inference costs.
Here, the function doubleDequant$(c_1^{\text{FP32}}, c_2^{\text{k-bit}}, \mathbf{W}^{\text{k-bit}})$ can be obtained by:
\begin{equation}
\text{dequant}(\text{dequant}(c_1^{\text{FP32}}, c_2^{\text{k-bit}}), \mathbf{W}^{\text{4bit}}) = \mathbf{W}^{\text{BF16}},
\end{equation}
The matrix $\mathbf{W}$ is of size NF4 and is quantized to 4-bit precision.
The constant $c_2$ is of size FP8.
We choose a blocksize of 64 for $\mathbf{W}$ to achieve higher quantization precision, and a blocksize of 256 for $c_2$ to conserve memory.
This choice is based on recent results showing that 8-bit quantization does not significantly degrade performance compared to 16-bit quantization. 

In summary, QLoRA employs a singular storage data type, typically NF4, alongside a computational data type represented by 16-bit BrainFloat (BF16). 
The forward and backward passes entail dequantizing the storage data type to the computational data type. 
Notably, during these processes, weight gradients are exclusively computed for the LoRA parameters utilizing the BF16 representation.

\subsection{Auto-Regressive Decoding}
\label{sec:decode}
Beam search constitutes a fundamental algorithm employed across numerous NLP models, serving as the pivotal decision-making layer responsible for selecting the optimal output concerning predetermined target variables, such as maximum probability or the subsequent output character. 
This method facilitates the simultaneous consideration of multiple tokens for a specific position within a given sequence, relying on conditional probability assessments. 
The selection process is governed by a key hyperparameter, denoted as beam width, which determines the number of N-best alternatives to be considered. 
In our auto-regressive decoding process of AutoMage, we implement beam search with a beam width set to 4. 
This strategic choice significantly elevates the precision and quality of text generation by enabling the model to explore a wider range of possibilities and select the most suitable output. 

\section{AutoMage2: An Upgraded Version of AutoMage}
\label{sec:automage2}

AutoMage, fine-tuned with datasets comprising approximately 1500 EDA tool instructions, may exhibit overfitting to these limited datasets. 
Furthermore, Llama2 exhibits deficiencies in coding and logical reasoning capabilities \cite{luo2023wizardcoder}.
Consequently, AutoMage inherits these limitations undoubtedly. 
Considering that real-world environments present greater challenges, there are still opportunities to further enhance AutoMage's abilities.
AutoMage2 is an upgraded version of AutoMage that aims to be a more stable and capable controller of ChatEDA. 
Our objective is to augment AutoMage's capabilities, particularly in logical reasoning for task decomposition and script generation coding, to enhance the system's overall performance.
We adopt the model architecture (Section \ref{sec:model-arc}), efficient fine-tuning process (Section \ref{sec:qlora}), and auto-regressive decoding (Section \ref{sec:decode}) from AutoMage, and employ three key techniques to improve its capabilities: enriched training corpus, instruction tuning with explanation \cite{mukherjee2023orca}, and chain of thoughts (CoT) prompting \cite{wei2022chain}.

\subsection{Enriched Training Corpus.} 
\label{sec:enrich_corpus}
AutoMage's capabilities stem largely from the pretraining of the underlying LLMs. 
To enhance performance, we must also enhance the base LLMs.

First, a high-quality corpus is critical for instruction tuning. 
With only around 1500 instances generated by the self instruction paradigm (Section \ref{sec:self-instruct}), AutoMage achieved high performance. 
To augment the data, we not only added more EDA tool usage instances but also filtered to ensure quality and semantic diversity. 
Specifically, we had an instructor \cite{su2023one} for encoding the generated instructions, calculated cosine similarity, and removed those with a similarity score above 0.95. 
After deduplication, we obtained around 1500 instances.

Moreover, code corpus represents a highly abstract language containing complex logical constructs. 
Exposure to such corpus can strengthen LLMs' coding skills and logical reasoning. 
To strategically bolster the coding and reasoning proficiencies of the LLMs, we amalgamated approximately 110k instances of code instructions \cite{luo2023wizardcoder} sourced from open-access repositories with our own generated instances of EDA tools instructions after deduplication. 
This amalgamation resulted in the creation of hybrid instruction datasets. These hybrid datasets, rich in diversity and complexity, were instrumental in fine-tuning the instructions provided to AutoMage2, thereby augmenting their overall comprehension and proficiency in utilizing EDA tools effectively. 
This amalgamation not only broadened the scope of the instructions but also fostered a comprehensive understanding of the intricate nuances associated with coding and EDA, ensuring a robust and well-rounded training process.

\subsection{Instruction Tuning with Explanation.}
The training corpus contains teacher (GPT-3.5/4) responses that explain the reasoning process, providing additional learning signals beyond the prompt-response pairs used in vanilla instruction tuning (Section \ref{sec:self-instruct}) for AutoMage.
To elicit such explanations, we employ system instructions for instruction tuning with the explanation of AutoMage2 following Orca \cite{mukherjee2023orca} style (e.g. ``think step-by-step'' and ``justify your steps''), which can better enable mimicking the teacher's thought process.
The intricately designed prompt is presented below for reference:
\begin{bluebox}[minted language=text, minted options={fontsize=\scriptsize}]{Instruction Fine-tuning Prompt}
### System:
You are AI assistant, capable of utilizing numerous tools and functions. User will give you a task. @\textbf{Your job is to generate a Python script to complete the task using the provided tools and functions.}@ While performing the task @\textbf{think step-by-step}@ and @\textbf{justify your steps}@.
You have access to the following tools and functions:
chateda is an autonomous tool that can automate the RTL to GDSII flow by executing steps through tools(functions) with various parameters.
tune is a function that can perform parameter tuning.
Specifically, you have access to the following details of the provided tools and functions:
<<<API documents>>>
### User:
<<<Requirement>>>
### Assistant:
<<<Response>>>
\end{bluebox}

\subsection{Chain of Thoughts.}
CoT prompting \cite{wei2022chain} enables complex reasoning through intermediate reasoning steps. 
Zero-shot CoT prompting \cite{kojima2022large} builds on this by introducing a simple zero-shot prompt: appending ``Let's think step by step.'' 
This prompts LLMs to generate a CoT that answers the question. From this, more accurate answers can be extracted.
Consequently, we apply zero-shot CoT during inference of AutoMage models and other notable LLMs to obtain more precise responses.
The intricately designed zero-shot CoT prompt is presented below for reference:
\begin{bluebox}[minted language=text, minted options={fontsize=\scriptsize}]{Zero-Shot CoT Prompt}
### System:
You are AI assistant, capable of utilizing numerous tools and functions. User will give you a task. Your job is to generate a Python script to complete the task using the provided tools and functions.
While performing the task think step-by-step and justify your steps.
You have access to the following tools and functions:
chateda is an autonomous tool that can automate the RTL to GDSII flow by executing steps through tools(functions) with various parameters.
tune is a function that can perform parameter tuning.
Specifically, you have access to the following details of the provided tools and functions:
<<<API documents>>>
### User:
<<<Requirement>>>
@\textbf{Let's first describe and explain what the task is asking. Then, analyze how to complete the task step by step using the provided tools and functions. Finally, generate the Python script according to your analysis.}@
### Assistant:
<<<Response>>>
\end{bluebox}

\section{Experiments}
\label{sec:exp}

\subsection{Setup}
\label{sec:setup}
For efficient fine-tuning of AutoMage2, we implement a constant learning rate schedule with a 0.03 warm-up ratio using paged AdamW 8-bit optimizer \cite{dettmers20218}, initiating with a learning rate of $1 \times 10^{-4}$, no weight decay, a batch size of 128, and a sequence length of 4096 tokens.
Ultimately, the model is fine-tuned for 1 epoch on 16$\times$A100 with 80G memory each.

During the inference phase, the user‘s requirement prompts in natural language.
These can be designed for a simple task (e.g. ``Perform routing for the processor design on the asap7 platform.'') or delineate a broader, more general goal (e.g. ``Please show me how to complete the design flow in the script.'').
The output of AutoMage2 is the executable script.

As for evaluation, we consider notable LLMs including Claude2 \cite{2023claude2}, GPT-3.5 \cite{brown2020language}, and GPT-4 \cite{openai2023gpt4} as our baselines for performance assessment.
To ensure a comprehensive comparison, we utilize different LLMs as the core controllers for our autonomous agent, ChatEDA.
The target API is a simplified Python wrapper of OpenROAD~\cite{ajayi2019openroad}.

For better understanding and easy reproduction of our work, we provide ChatEDA-bench (\Cref{sec:ChatEDA-bench}), examples of EDA tool instructions dataset, and API document with its corresponding OpenROAD implementation in the open-source repo \url{https://github.com/wuhy68/ChatEDAv1}.

\subsection{ChatEDA-Bench}
\label{sec:ChatEDA-bench}
To assess the effectiveness of AutoMage2, we have developed ChatEDA-Bench, a comprehensive evaluation benchmark comprising 50 distinct tasks spanning three distinct categories: simple flow calls (30\%), complex flow calls (30\%), and parameter flow calls (40\%). The diversity of these tasks ensures a rigorous evaluation of AutoMage2's capabilities across various scenarios. In the subsequent sections, we will elucidate two examples from each task category, shedding light on the intricate nature of the evaluation scenarios and the challenges posed by each case. 

\minisection{Simple Flow Calls}
The first task requires the successful execution of the whole process, including evaluation. These cases test the fundamental application of LLMs and their sequence of usage with the API interfaces. 
\begin{minted}[frame=lines,framerule=2pt,linenos=false,fontsize=\scriptsize]{text}
Case 1: I want to carry out clock tree synthesis for the design "modulator.v" using the process design kit "gf180" with a desired density of 0.95. Can you help me to generate the script?

Case 2: For the "aes" circuit, I want to run the steps from setup to detailed routing on the platform "asap7"?  
\end{minted}

\minisection{Complex Flow Calls}
These cases heavily rely on logic, including traversing parameters, further examining the LLM's logical reasoning and understanding of each API argument.
\begin{minted}[frame=lines,framerule=2pt,linenos=false,fontsize=\scriptsize]{text}
Case 1: I want to perform a grid search on the design "how" on the "gf180" platform for floorplan parameters, CTS parameters, and placement parameters to find the best balance of chip area, power consumption and performance. Can you help me to do that?

Case 2: I have a design called "data_processor" on the platform "asap7". I want to experiment with different combinations of clock periods, density values, halo sizes, and channel widths during the whole EDA flow.
\end{minted}

\minisection{Parameter Tuner Calls}
These cases require the LLM to provide a parameter-tuning solution, thoroughly testing the LLM's logic and use of EDA tools.
\begin{minted}[frame=lines,framerule=2pt,linenos=false,fontsize=\scriptsize]{text}
Case 1: I want to perform DSE for the design "aaksdjka" using the technology node "asap7", and consider tns, wns, the required chip surface area, and the power consumption at the final stage as the evaluation metric of DSE. The search space of parameters are as follows:
1. Core utilization ranging from 60% to 80% with a step of 5%.
2. Core aspect ratio from 1 to 3 continuously.
3. Core margins from 2 to 10 with a step of 1.
4. Macro place halo from 5 to 15 with a step of 1.
5. Macro place channel from 5 to 15 with a step of 1.
6. Clock period of 5.
7. Density from 0.3 to 0.9 with a step of 0.
8. 50% TNS end percent.

Case 2: I want to perform DSE for the design "aaksdjka" using the technology node "asap7", and consider tns, wns, the required chip surface area, and the power consumption at the final stage as the evaluation metric of DSE. The search space of parameters includes placement parameters and clock tree synthesis parameters.
\end{minted}

\subsection{Evaluation of LLMs}
\label{sec:evaluation}
An objective evaluation system was developed to assess the task decomposition and script generation capabilities of LLMs designed for automated script generation.
An evaluation system was developed to assess the task decomposition and script generation capabilities of LLMs designed for automated script generation.
This evaluation system comprises two integral components. 
Initially, the Python scripts generated are subjected to testing through the EDA tool interface to determine their executability. 
Subsequently, a manual assessment is conducted to ascertain whether the responses from the LLMs meet the users' requirements. 
To maintain fairness and accuracy in the evaluation process, multiple judges' perspectives are taken into account, and these judges are kept unaware of which LLM generated the response during scoring.
The system uses a three-tiered grading scheme, with Grade A representing the highest achievement.
Grade A is awarded to LLMs that demonstrate coherent task decomposition and generate accurate scripts.
Grade B indicates respectable but imperfect performance, assigned to LLMs that plan logically but falter in script generation.
Grade C denotes failure in both task decomposition and code generation. 

\begin{figure}[]
    \centering
    \includegraphics[width=.80\linewidth]{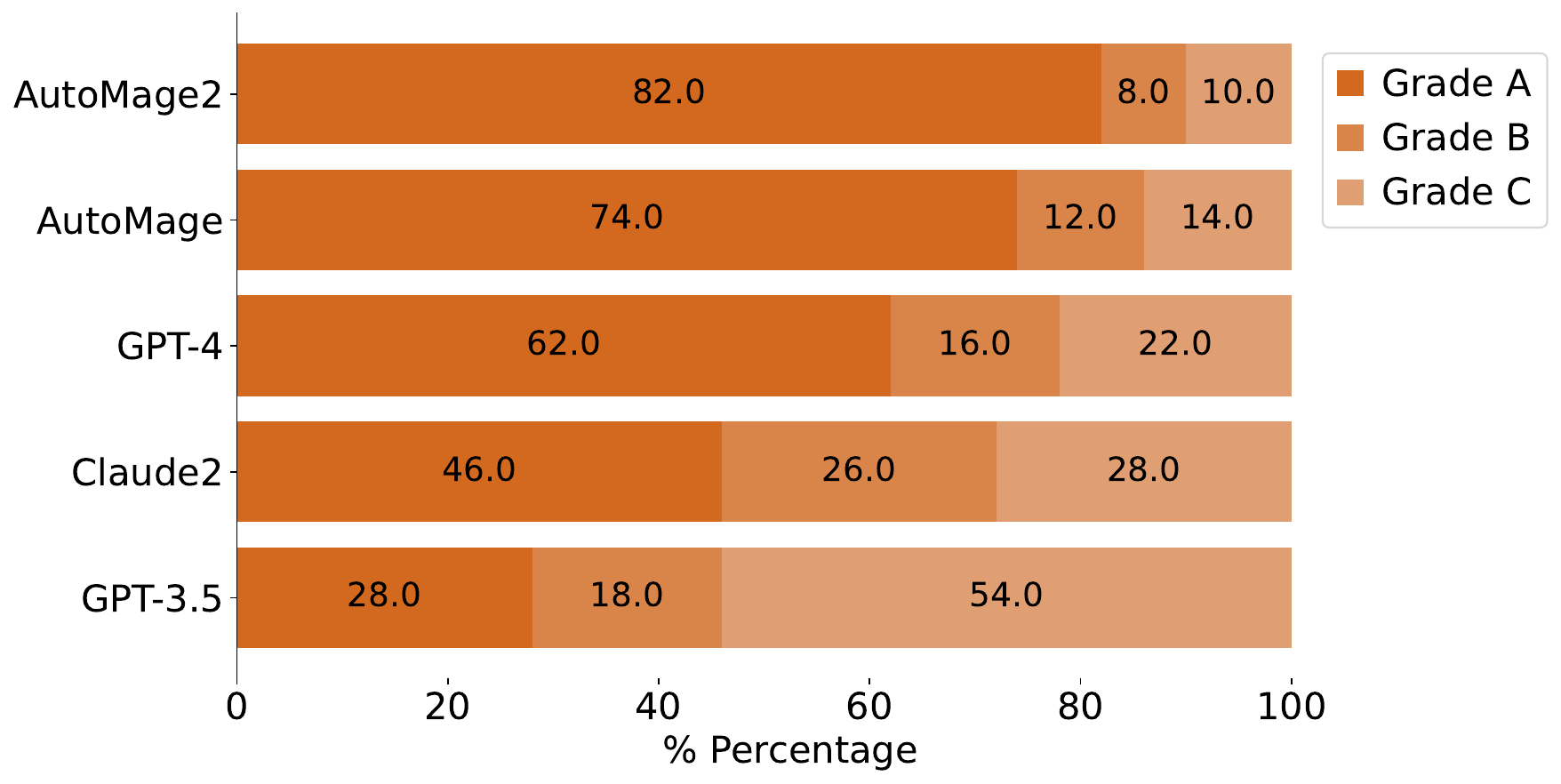} 
    \caption{Evaluation results for AutoMage \cite{he2023chateda} and AutoMage2 compared to other LLMs. 
    AutoMage models outperform other notable LLMs by a significant margin in task planning and script generation and AutoMage2 performs the best.}
    \label{figs:script_evaluation}
\end{figure}

During the evaluation process, we use ChatEDA-Bench for a comprehensive inspection.
As summarized in \Cref{figs:script_evaluation}, AutoMage models outperform all notable LLMs, and our proposed AutoMage2 achieved the best performance, correctly earning Grade A for 82\% of test cases.
This significantly exceeds the 62\% Grade A attained by the next highest performer of notable LLMs, GPT-4.
While GPT-4 exhibited reasonably strong capabilities, it struggled to differentiate between lower quality responses, assigning Grade B and C more evenly at 16\% and 22\% respectively.
In contrast, AutoMage2 reliably identified the highest grade-worthy responses.
The other LLMs, Claude2 and GPT-3.5, performed noticeably worse than AutoMage2, with Claude2 achieving 46\% Grade A and GPT-3.5 only 28\% Grade A.
This highlights their comparative deficiencies.
In summary, AutoMage2 substantially outperformed GPT-4 and other LLMs by consistently assigning appropriate high grades.
The considerable margin between AutoMage2 and second-place GPT-4 demonstrates the advances of our method in grading performance. 
Our experiment affirms AutoMage2 as the current state-of-the-art LLM for this evaluation task, which will be the most reliable controller of ChatEDA.

\subsection{Case Studies}
\label{sec:case_studies}
In this section, we analyze five tasks involving user interactions with EDA tools through a conversational interface.
It is important to note that this exploration is not exhaustive due to the limitations inherent in this qualitative study.
For instance, controlling all possible variables, such as the vast array of responses the LLM can generate for a given prompt, is beyond our reach.
As a result, our conclusions are based on a selected group of samples deemed representative.
Nonetheless, we contend that the description of these exemplars will enhance our understanding of the robust performance of AutoMage2, which enhances the reliability of AutoMage2 powered ChatEDA.
For each task, we first outline the user's requirements and subsequently present the task decomposition process and the scripts generated by AutoMage2.
Moreover, given that GPT-4 is the best LLM in baselines during the script evaluation, we also provide the scripts generated by GPT-4 for comparison.

\minisection{Task 1: Parameter Grid Search}.
The first task requires LLMs to execute a grid search, heavily relying on logic like traversing parameters. 
This further examines the LLM's grasp of each API argument. 
AutoMage2 performed excellently, demonstrating its logical prowess and exemplary API argument comprehension. 
In contrast, GPT-4 showed a deficient understanding of parameter combination optimization.
This highlights AutoMage2's more sophisticated grasp of the EDA tools' interface and parameter tuning logic compared to GPT-4.
\begin{itemize}[leftmargin=0cm, itemindent=0cm]         
    \item[] \textbf{\textsc{User:}} I want to perform a grid search on the design ``how'' on ``gf180'' platform for floorplan parameters, CTS parameters and placement parameters to find the best balance of chip area, power consumption and performance. Can you help me to do that?

    \hdashrule{\columnwidth}{1pt}{3pt}
    \item[] \textbf{\textsc{AutoMage2:}} 
    \begin{itemize}[leftmargin=0cm, itemindent=0cm]
        \item[] 
\begin{purplebox}[minted language=text]{Task Decomposition}
This task requires us to perform a grid search over a set of parameters for the floorplan, CTS, and placement stages of the chip design flow. The parameters to be tuned are:
1. Floorplan: core_utilization, core_aspect_ratio, core_margins, macro_place_halo, macro_place_channel
2. CTS: tns_end_percent
3. Placement: density

The user wants to find the best balance between chip area, power consumption, and performance. Therefore, we need to get the "area", "power", and "performance" metrics after the final report stage.

Here is the step-by-step analysis of how to complete this task:
1. Initialize the EDA tool using the "setup" function with the design name "how" and the platform "gf180".
2. Run logic synthesis using the "run_synthesis" function.
3. Perform floorplan using the "floorplan" function, iterating over the values of the parameters to be tuned.
4. Perform placement using the "placement" function, iterating over the values of the parameters to be tuned.
5. Perform CTS using the "cts" function, iterating over the values of the parameters to be tuned.
6. Perform global routing using the "global_route" function.
7. Perform detailed routing using the "detail_route" function.
8. Perform density fill using the "density_fill" function.
9. Generate the final report using the "final_report" function.
10. Get the "area", "power", and "performance" metrics using the "get_metric" function.
\end{purplebox}
        \item[]
\begin{purplebox}[minted language=python]{Script Generation}
# Define the parameter space for the grid search
floorplan_parameters = [
    {"core_utilization": 70, "core_aspect_ratio": 1, "core_margins": 10, "macro_place_halo": 5, "macro_place_channel": 10},
    {"core_utilization": 80, "core_aspect_ratio": 1.2, "core_margins": 15, "macro_place_halo": 10, "macro_place_channel": 15},
    {"core_utilization": 90, "core_aspect_ratio": 1.5, "core_margins": 20, "macro_place_halo": 15, "macro_place_channel": 20},
]
placement_parameters = [
    {"density": 0.6},
    {"density": 0.7},
    {"density": 0.8},
]
cts_parameters = [
    {"tns_end_percent": 20},
    {"tns_end_percent": 30},
    {"tns_end_percent": 40},
]

# Perform the grid search
best_parameters = {"floorplan": None, "placement": None, "cts": None}
best_performance = float("inf")
for floorplan_parameter in floorplan_parameters:
    for placement_parameter in placement_parameters:
        for cts_parameter in cts_parameters:
            eda = chateda()
            eda.setup(design_name="how", platform="gf180")
            eda.run_synthesis()
            eda.floorplan(
                core_utilization=floorplan_parameter["core_utilization"],
                core_aspect_ratio=floorplan_parameter["core_aspect_ratio"], 
                core_margins=floorplan_parameter["core_margins"], 
                macro_place_halo=floorplan_parameter["macro_place_halo"], 
                macro_place_channel=floorplan_parameter["macro_place_channel"]
            )
            eda.placement(density=placement_parameter["density"])
            eda.cts(tns_end_percent=cts_parameter["tns_end_percent"])
            eda.global_route()
            eda.detail_route()
            eda.density_fill()
            eda.final_report()
            performance = eda.get_metric("final", ["area", "power", "performance"])
            if performance < best_performance:
                best_performance = performance
                best_parameters["floorplan"] = floorplan_parameter
                best_parameters["placement"] = placement_parameter
                best_parameters["cts"] = cts_parameter
print(best_parameters)
\end{purplebox}
    \end{itemize}

    \hdashrule{\columnwidth}{1pt}{3pt}
    \item[] \textbf{\textsc{GPT-4:}} 
    \begin{itemize}[leftmargin=0cm, itemindent=0cm]
        \item[] 
\begin{greenbox}[minted language=python]{Script Generation}
# Instantiate the chateda class
eda = chateda()
# Setup the EDA tool for the "how" design on "gf180" platform
eda.setup(design_name="how", platform="gf180")
# Define the floorplan function for grid search
def floorplan_func(core_utilization, core_aspect_ratio, core_margins):
    eda.floorplan(core_utilization=core_utilization, core_aspect_ratio=core_aspect_ratio, core_margins=core_margins)
    return eda.get_metric(stage="floorplan", metrics=["area", "power", "performance"])
# Define the CTS function for grid search
def cts_func(tns_end_percent):
    eda.cts(tns_end_percent=tns_end_percent)
    return eda.get_metric(stage="cts", metrics=["area", "power", "performance"])
# Define the placement function for grid search
def placement_func(density):
    eda.placement(density=density)
    return eda.get_metric(stage="place", metrics=["area", "power", "performance"])

# Grid search parameters for each stage
floorplan_params = {
    "core_utilization": {"minmax": [50, 100], "step": 10},
    "core_aspect_ratio": {"minmax": [0.5, 2.0], "step": 0.5},
    "core_margins": {"minmax": [1, 5], "step": 1}
}
cts_params = {
    "tns_end_percent": {"minmax": [10, 100], "step": 10}
}
placement_params = {
    "density": {"minmax": [0.1, 1.0], "step": 0.1}
}

# Perform grid search
tune(floorplan_func, floorplan_params)
tune(cts_func, cts_params)
tune(placement_func, placement_params)
\end{greenbox}
    \end{itemize}
\end{itemize}

\minisection{Task 2: Parameter Tuning Process}.
The next task requires the model to provide a parameter-tuning solution. 
Examining the LLM's use of the tune function also evaluates its ability to suggest an appropriate parameter space. 
Without sufficient priors, LLMs cannot determine an exact parameter range. 
However, we can see that both AutoMage2 and GPT-4 can generate a complex search space, endowing the parameter tuning strategy with greater potential. 
However, GPT-4 misses an important intermediate "synthesis" step in task decomposition, demonstrating weaker comprehension of the task decomposition. 
In this aspect, AutoMage2 excels.
\begin{itemize}[leftmargin=0cm, itemindent=0cm]         
    \item[] \textbf{\textsc{User:}} What are the tunable parameters in floorplan? Give me an example of tuning them for "router" design on the platform sky130 to get the ppa results. Try to provide a larger search space for parameter tuning.

    \hdashrule{\columnwidth}{1pt}{3pt}
    \item[] \textbf{\textsc{AutoMage2:}} 
    \begin{itemize}[leftmargin=0cm, itemindent=0cm]
        \item[] 
\begin{purplebox}[minted language=text]{Task Decomposition}
To complete this task, we need to first identify the tunable parameters in the floorplan function. According to the chateda API documentation, the tunable parameters in the floorplan function are core_utilization, core_aspect_ratio, core_margins, macro_place_halo, and macro_place_channel.

Next, we need to write a function that takes these parameters as inputs, runs the flow using these parameters and returns the ppa results.

Finally, we can use the tune function to tune these parameters. We need to provide a dictionary where the keys are the parameter names and the values are dictionaries containing the ranges and step sizes for these parameters.
\end{purplebox}
        \item[] 
\begin{purplebox}[minted language=python]{Script Generation}
def tune_floorplan(core_utilization, core_aspect_ratio, core_margins, macro_place_halo, macro_place_channel):
    eda_tool = chateda()
    eda_tool.setup(design_name="router", platform="sky130")
    eda_tool.run_synthesis()
    eda_tool.floorplan(core_utilization=core_utilization, core_aspect_ratio=core_aspect_ratio, core_margins=core_margins, macro_place_halo=macro_place_halo, macro_place_channel=macro_place_channel)
    eda_tool.placement()
    eda_tool.cts()
    eda_tool.global_route()
    eda_tool.detail_route()
    eda_tool.density_fill()
    eda_tool.final_report()
    return eda_tool.get_metric("final", ["power", "performance", "area"])
    
tunable_params = {
    "core_utilization": {"minmax": [0, 100], "step": 5},
    "core_aspect_ratio": {"minmax": [0.5, 2], "step": 0.1},
    "core_margins": {"minmax": [5, 20], "step": 1},
    "macro_place_halo": {"minmax": [5, 20], "step": 1},
    "macro_place_channel": {"minmax": [5, 20], "step": 1}
}
tune(tune_floorplan, tunable_params)
\end{purplebox}
    \end{itemize}
    \hdashrule{\columnwidth}{1pt}{3pt}
    \item[] \textbf{\textsc{GPT-4:}} 
    \begin{itemize}[leftmargin=0cm, itemindent=0cm]
        \item[] 
\begin{greenbox}[minted language=python]{Script Generation}
# Define the target function for parameter tuning
def target_function(core_utilization, core_aspect_ratio, core_margins, macro_place_halo, macro_place_channel
    # Initialize chateda class
    eda_tool = chateda()
    # Setup EDA tool with design_name as "router" and platform as "sky130"
    eda_tool.setup(design_name="router", platform="sky130")
    eda_tool.floorplan(
        core_utilization=core_utilization,
        core_aspect_ratio=core_aspect_ratio,
        core_margins=core_margins,
        macro_place_halo=macro_place_halo,
        macro_place_channel=macro_place_channel
    )
    return eda_tool.get_metric(stage="floorplan", metrics=["power", "performance", "area"])
    
# Parameters to be tuned with their respective min, max, and step values
params = {
    "core_utilization": {"minmax": [50, 100], "step": 5},
    "core_aspect_ratio": {"minmax": [0.5, 2], "step": 0.1},
    "core_margins": {"minmax": [1, 5], "step": 1},
    "macro_place_halo": {"minmax": [1, 10], "step": 1},
    "macro_place_channel": {"minmax": [1, 10], "step": 1}
}
# Tune parameters
tune(target_function, params)  
\end{greenbox}
    \end{itemize}
\end{itemize}

\minisection{Task 3: Customized Optimization}.
Customized Optimization requires tuning parameters in the model per the search space specified by the user. 
This process thoroughly evaluates the model's understanding of the tune function and the significance of each API interface parameter.
It is clear that AutoMage2 appropriately grasps the need for user-customized parameter tuning, while GPT-4 exhibits a deficient understanding of key EDA tool arguments. 
This demonstrates AutoMage2's more sophisticated comprehension of the EDA tools' interface and the logic of parameter tuning compared to GPT-4.
\begin{itemize}[leftmargin=0cm, itemindent=0cm]         
    \item[] \textbf{\textsc{User:}} We are going to develop the new ASIC project ``datacenter\_chip'' for server applications on the ``nangate4'' platform, I want to create an automated script that will tune the EDA process, while optimizing for minimal power usage and maintaining acceptable performance and area. Here is my plan:
    \begin{enumerate}
    \item Density level ranges from 0.1 to 0.9 with increments of 0.005;
    \item The search space of core utilization percentages is between 60\% and 85\%;
    \item Clock periods are ranging from 5 to 10 in steps of 1;
    \item Resolve 50\% to 80\% of timing violations, increasing in steps of 3\%;
    \item Setting the core aspect ratio play around between 1 and 2 with a step of 0.2;
    \item Setting the core margins, macro place halo, and macro place channel to be 7, 8, 9 separately;
    \item Keeping the rest of the parameters at their default values.
    \end{enumerate}

    \hdashrule{\columnwidth}{1pt}{3pt}
    \item[] \textbf{\textsc{AutoMage2:}} 
    \begin{itemize}[leftmargin=0cm, itemindent=0cm]
        \item[] 
\begin{purplebox}[minted language=text]{Task Decomposition}
This task requires us to develop a script that automates the EDA process for the "datacenter_chip" project on the "nangate4" platform. The goal is to optimize the design for minimal power usage while maintaining acceptable performance and area.

To achieve this, we need to tune several parameters of the EDA process, including density, core utilization, clock period, percentage of timing violations to resolve, core aspect ratio, core margins, macro place halo, and macro place channel.

Here is a step-by-step analysis of how to complete the task:
1. Initialize the EDA tool with the design name "datacenter_chip" and platform "nangate4".
2. Run logic synthesis with varying clock periods.
3. Perform floorplanning with different core utilization percentages and core aspect ratios.
4. Run placement with different densities.
5. Run clock tree synthesis with different percentages of timing violations to resolve.
6. Run global routing and detail routing with default parameters.
7. Run density fill and final report with default parameters.
8. Evaluate the performance metrics for power, performance, and area.
\end{purplebox}
        \item[] 
\begin{purplebox}[minted language=python]{Script Generation}
def tune_flow(density, core_utilization, clock_period, tns_end_percent, core_aspect_ratio, core_margins, macro_place_halo, macro_place_channel):
    eda = chateda()
    eda.setup(design_name="datacenter_chip", platform="nangate4")
    eda.run_synthesis(clock_period=clock_period)
    eda.floorplan(core_utilization=core_utilization, core_aspect_ratio=core_aspect_ratio, core_margins=core_margins, macro_place_halo=macro_place_halo, macro_place_channel=macro_place_channel)
    eda.placement(density=density)
    eda.cts(tns_end_percent=tns_end_percent)
    eda.global_route()
    eda.detail_route()
    eda.density_fill()
    eda.final_report()
    return eda.get_metric("final", ["power", "performance", "area"])
    
params = {
    "density": {"minmax": [0.1, 0.9], "step": 0.005},
    "core_utilization": {"minmax": [60, 85], "step": 1},
    "clock_period": {"minmax": [5, 10], "step": 1},
    "tns_end_percent": {"minmax": [50, 80], "step": 3},
    "core_aspect_ratio": {"minmax": [1, 2], "step": 0.2},
    "core_margins": {"minmax": [7, 7], "step": 0},
    "macro_place_halo": {"minmax": [8, 8], "step": 0},
    "macro_place_channel": {"minmax": [9, 9], "step": 0}
}
tune(tune_flow, params)
\end{purplebox}
    \end{itemize}

    \hdashrule{\columnwidth}{1pt}{3pt}
    \item[] \textbf{\textsc{GPT-4:}} 
    \begin{itemize}[leftmargin=0cm, itemindent=0cm]
        \item[] 
\begin{greenbox}[minted language=python]{Script Generation}
def run_eda_flow(density, core_util, clock_period, tns_end_percent, core_aspect_ratio, core_margin, macro_halo, macro_channel):
    # Setup the design
    tool.setup(design_name="datacenter_chip", platform="nangate4")
    # Run synthesis with the given clock period
    tool.run_synthesis(clock_period=clock_period)
    # Perform floorplanning with the specified parameters
    tool.floorplan(core_utilization=core_util, core_aspect_ratio=core_aspect_ratio, core_margins=core_margin, macro_place_halo=macro_halo, macro_place_channel=macro_channel)
    # Run placement with the specified density
    tool.placement(density=density)
    # Execute CTS to address the specified percentage of timing violations
    tool.cts(tns_end_percent=tns_end_percent)
    # Continue with the default parameters for the subsequent EDA steps
    tool.global_route()
    tool.detail_route()
    tool.density_fill()
    tool.final_report()
    # Retrieve the power, performance, and area metrics after the EDA flow
    power_metric = tool.get_metric(stage="final", metrics=["power"])
    performance_metric = tool.get_metric(stage="final", metrics=["performance"])
    area_metric = tool.get_metric(stage="final", metrics=["area"])
    # The goal is to minimize power consumption, so we'll use power as our primary metric
    return power_metric
    
# Parameter space as defined by the user
param_space = {
    "density": {"minmax": [0.1, 0.9], "step": 0.005},
    "core_util": {"minmax": [60, 85], "step": 1},
    "clock_period": {"minmax": [5, 10], "step": 1},
    "tns_end_percent": {"minmax": [50, 80], "step": 3},
    "core_aspect_ratio": {"minmax": [1, 2], "step": 0.2},
    "core_margin": {"minmax": [7, 9], "step": 1},
    "macro_halo": {"minmax": [7, 9], "step": 1},
    "macro_channel": {"minmax": [7, 9], "step": 1}
}
tune(func=run_eda_flow, param=param_space)
\end{greenbox}
    \end{itemize}
\end{itemize}

\minisection{Task 4: Clock Period Minimization}.
During instruction tuning, we guided AutoMage2 on using the API interface according to our specific needs. 
However, we did not teach it to modify the parameters using self-feedback from the evaluation results.
In this test case, we examined how the LLM adjusts the API arguments through self-feedback to achieve the user-requested evaluation outcomes. 
Intriguingly, both AutoMage2 and GPT-4 successfully completed the provided test case.
\begin{itemize}[leftmargin=0cm, itemindent=0cm]         
    \item[] \textbf{\textsc{User:}} Try to find out the smallest valid clock period for the design ``leon'' on ``asap7'' platform. Note that a clock period is valid only if the ``wns'' metric at the final stage is non negative.
    \hdashrule{\columnwidth}{1pt}{3pt}
    \item[] \textbf{\textsc{AutoMage2:}} 
    \begin{itemize}[leftmargin=0cm, itemindent=0cm]
        \item[] 
\begin{purplebox}[minted language=text]{Task Decomposition}
To achieve this, we can follow these steps:
1. Set up the EDA tool with the design name "leon" and platform "asap7".
2. Run logic synthesis with the smallest possible clock period (default value).
3. Perform floorplanning, placement, CTS, global routing, detail routing, density fill, and final report.
4. Get the "wns" metric after the final stage.
5. If the "wns" metric is non-negative, return the current clock period. Otherwise, increase the clock period and repeat the process.
\end{purplebox}
        \item[] 
\begin{purplebox}[minted language=python]{Script Generation}
def find_smallest_valid_clock_period(clock_period):
    # Create an instance of the chateda class
    eda_tool = chateda()
    # Setup the design
    eda_tool.setup(design_name="leon", platform="asap7")
    # Run logic synthesis
    eda_tool.run_synthesis(clock_period=clock_period)
    # Run floorplanning
    eda_tool.floorplan()
    # Run placement
    eda_tool.placement()
    # Run CTS
    eda_tool.cts()
    # Run global routing
    eda_tool.global_route()
    # Run detail routing
    eda_tool.detail_route()
    # Run density fill
    eda_tool.density_fill()
    # Run final report
    eda_tool.final_report()
    # Get the "wns" metric
    wns = eda_tool.get_metric("final", ["wns"])
    if wns >= 0:
        return clock_period
    else:
        return find_smallest_valid_clock_period(clock_period+1)
# Call the function to find the smallest valid clock period
smallest_valid_clock_period = find_smallest_valid_clock_period(1)
print(smallest_valid_clock_period)
\end{purplebox}
    \end{itemize}
    \hdashrule{\columnwidth}{1pt}{3pt}
    \item[] \textbf{\textsc{GPT-4:}} 
    \begin{itemize}[leftmargin=0cm, itemindent=0cm]
        \item[] 
\begin{greenbox}[minted language=python]{Script Generation}
# Starting from an assumed clock period. This might need tweaking based on prior knowledge of the design and platform.
clock_period = 1.0  # Initial assumption in nanoseconds (ns)
while True:
    # Initializing the chateda class
    eda_tool = chateda()
    # Setup for the 'leon' design on 'asap7' platform
    eda_tool.setup(design_name="leon", platform="asap7")
    # Running the synthesis with the current clock_period
    eda_tool.run_synthesis(clock_period=clock_period)
    # Following the EDA flow
    eda_tool.floorplan()
    eda_tool.placement()
    eda_tool.cts()
    eda_tool.global_route()
    eda_tool.detail_route()
    eda_tool.density_fill()
    eda_tool.final_report()
    # Fetching the WNS metric at the final stage
    wns = eda_tool.get_metric(stage="final", metrics=["wns"])
    # If WNS is non-negative, then the current clock_period is valid
    if wns >= 0:
        break
    else:
        clock_period += 0.1  # Incrementing the clock period by 0.1ns. This step can also be tweaked.
print(f"The smallest valid clock period for the design 'leon' on 'asap7' platform is {clock_period} ns.")
\end{greenbox}
    \end{itemize}
\end{itemize}

\minisection{Task 5: Self Correct}.
This task evaluates the model's ability to self-correct mistakes, while also testing its capacity to comprehend EDA tools from the opposite perspective. 
Both AutoMage2 and GPT-4 can accurately identify and fix errors, demonstrating AutoMage2's self-correction skills and profound understanding of utilizing EDA tools.
\begin{itemize}[leftmargin=0cm, itemindent=0cm]         
    \item[] \textbf{\textsc{User:}} Can you help me check whether the provided script can complete the task and provide the reason? 
    
    Here is the task:
    
    I want to optimize performance at the routing stage for my design asjdk on nangate45. 
    
    Here is the script:
\begin{minted}[frame=lines,framerule=1pt,linenos=false]{python}
def tune_performance():
    params = {"tns_end_percent": {"minmax": [0, 100], "step": 5}}
    def optimize(tns_end_percent):
        eda = chateda()
        eda.setup(design_name="asjdk", platform="nangate45")
        eda.run_synthesis()
        eda.cts(tns_end_percent=tns_end_percent)
        eda.global_route()
        eda.detail_route()
        return eda.get_metric(stage="route", metrics=["performance"])
    best_params = tune(optimize, params)
    print(best_params)
tune_performance()
\end{minted}
    If the script can't complete the task, please provide a script that can complete the task.
    
    \hdashrule{\columnwidth}{1pt}{3pt}
    
    \item[] \textbf{\textsc{AutoMage2:}} 
    \begin{itemize}[leftmargin=0cm, itemindent=0cm]
        \item[] 
        \begin{purplebox}[minted language=text]{Task Decomposition}
The task is to optimize the performance at the routing stage for the design "asjdk" on the "nangate45" platform. The script provided is almost correct, but it lacks some necessary steps.

Here are the steps to complete the task:
1. Set up the EDA tool with the design name "asjdk" and the platform "nangate45".
2. Run logic synthesis.
3. Perform floorplanning.
4. Run placement.
5. Perform clock tree synthesis (CTS) with different percentages of violating paths to fix.
6. Run global routing.
7. Run detail routing.
8. Get the performance metric after detail routing.
9. Tune the parameter "tns_end_percent" to find the value that gives the best performance.
\end{purplebox}
        \item[] 
\begin{purplebox}[minted language=python]{Script Generation}
def tune_performance():
    params = {"tns_end_percent": {"minmax": [0, 100], "step": 5}}
    def optimize(tns_end_percent):
        eda = chateda()
        eda.setup(design_name="asjdk", platform="nangate45")
        eda.run_synthesis()
        eda.floorplan()
        eda.placement()
        eda.cts(tns_end_percent=tns_end_percent)
        eda.global_route()
        eda.detail_route()
        return eda.get_metric(stage="route", metrics=["performance"])
    best_params = tune(optimize, params)
    print(best_params)
tune_performance()
\end{purplebox}
    \end{itemize}
    \hdashrule{\columnwidth}{1pt}{3pt}
    \item[] \textbf{\textsc{GPT-4:}} 
    \begin{itemize}[leftmargin=0cm, itemindent=0cm]
        \item[] 
\begin{greenbox}[minted language=python]{Script Generation}
def tune_performance():
    params = {"tns_end_percent": {"minmax": [0, 100], "step": 5}}
    def optimize(tns_end_percent):
        eda = chateda()
        eda.setup(design_name="asjdk", platform="nangate45")
        eda.run_synthesis()
        eda.floorplan()  # Adding the floorplan step.
        eda.placement()  # Adding the placement step.
        eda.cts(tns_end_percent=tns_end_percent)
        eda.global_route()
        eda.detail_route()
        return eda.get_metric(stage="route", metrics=["performance"])
    best_params = tune(optimize, params)
    print(best_params)
tune_performance()
\end{greenbox}
    \end{itemize}
\end{itemize}

\section{Discussion}
\label{sec:discussion}

Our proposed ChatEDA and its controllers, AutoMage models, achieve significant performance in the automation of the EDA flow. 
Despite the advancements, they face several limitations. 
Firstly, ChatEDA's design lacks universal applicability to diverse API documents. 
This constraint necessitates the creation of specialized API documentation for different EDA tools, thereby limiting ChatEDA's scalability and broader application in the EDA industry. 
Moreover, the unique API structure of each EDA tool requires a bespoke approach, impeding the seamless integration of ChatEDA across various platforms and tools.
Secondly, although AutoMage models demonstrate enhanced capability in EDA flow automation, they encompass a substantial number of model parameters. 
This complexity results in a significant slowdown during a single decoding step in these larger models. 
Thirdly, we haven't provided some basic evaluation flow or (semi-)automated scoring tools to support comparisons using the ChatEDA-Bench, which cannot enable a straightforward evaluation of different solutions from others.
Nevertheless, the assessment of LLM outputs presents challenges due to the vast output space, the subjective nature of relevance and coherence judgments, and considerations regarding human readability. 
Therefore, we have chosen to leverage human-centric evaluation with the input and insights of multiple expert judges, aiming to comprehensively evaluate not only the efficacy of task decomposition but also the quality of script generation within the context of LLM outputs.
We hope that the introduction of a benchmark comprising tasks remains highly beneficial for model debugging and comparative analysis of methodologies.

These limitations highlight the imperative for ongoing efforts to augment the versatility and efficiency of ChatEDA. 
Firstly, optimizing the decoding strategy of LLMs to improve their decoding speed would significantly enhance the usability and effectiveness of ChatEDA.
Moreover, a critical goal is to substantially enhance the generalization capacity of LLMs, enabling them to adeptly manage unfamiliar EDA tool documentation in zero-shot or few-shot scenarios. 
Such a breakthrough would significantly broaden the scope of LLMs in various EDA contexts, representing a major stride in AI-powered EDA tool integration and leading to more adaptable and versatile design automation solutions.
Additionally, we plan to advance the multi-turn dialogue capabilities of our system. This enhancement will facilitate dynamic interactions based on user feedback, allowing the system to amend and rectify potential errors in responses.

\section{Conclusion} 
\label{sec:conclu}

Interfacing EDA tools is essential for unleashing circuit design productivity.
In this work, we propose an LLM-powered autonomous agent for EDA, which enables a conversational interface for designers to interact with the design flow.
Technically, ChatEDA integrates a fine-tuned AutoMage, which orchestrates the design flow through task decomposition, script generation, and task execution.
ChatEDA handles various user requirements well, outperforming other LLM models like GPT-4 and so on.
We hope this work could inspire next-generation EDA tool evolution.





{
    \bibliographystyle{IEEEtran}
    \bibliography{ref/Top-sim,ref/FPGA,ref/FPGA-DNN,ref/DL,ref/ML,ref/RL,ref/SPEED,ref/ASIC,ref/bak-fpga,ref/Analog,ref/opc}
}

\end{document}